\documentclass[a4paper,11pt]{article}
\usepackage{pos}
\usepackage{float}
\usepackage[symbol]{footmisc}
\renewcommand{\thefootnote}{\fnsymbol{footnote}}
\newcommand{\KK}{${\cal KK}$}
\def\rQCED{{\rm QCED}}
\newcommand{\qb}{{\bar{q}}}
\newcommand{\sfac}{\mathfrak{s}}
\title{Recent Developments in IR-Improved Amplitude-Based Resummation in Precision High Energy Collider Physics}
\ShortTitle{Recent Developments in IR-Improved Amplitude-Based Resummation...}

\author*[a]{B.F.L. Ward}
\author[b]{S. Jadach\footnote[2]{Deceased.}}
\author[c]{W. Placzek}
\author[b]{M. Skrzypek}
\author[b]{Z.A. Was}
\author[d]{S.A. Yost}
\author[c]{A. Siodmok}

\affiliation[a]{Department of Physics, Baylor University,\\
 One Bear Place \# 97316, Waco, TX 76798-7316, USA}

\affiliation[b]{Institute of Nuclear Physics, Polish Academy of Sciences,\\
ul. Radzikowskiego 152, 31-342 Krakow, Poland}

\affiliation[c]{Institute of Applied Computer Science, Jagiellonian University,\\
ul. Prof. Stanisława Lojasiewicza 11, 30-348 Krakow, Poland}

\affiliation[d]{Department of Physics, The Citadel,\\
171 Moultrie Street, Charleston, SC 29409, USA}

\emailAdd{bfl\_ward@baylor.edu}

\abstract{We present recent developments in precision high energy collider physics based on the IR-improvement of unintegrable singularities in the infrared regime via amplitude-based resummation in $QED\times QCD \subset SU(2)_L \times U_1 \times SU(3)^c$. We focus on specific applications relevant to precision observables in LHC/FCC, LC, CLIC, CEPC, and CPPC physics, for which we present new results and some new issues.}
\centerline{BU-HEPP-26-01, Feb., 2026}
\FullConference{17th International Symposium on Radiative Corrections: Applications of Quantum Field Theory to Phenomenology (
RADCOR2025)\\
5th Oct. - 10th Oct., 2025\\
Puri, India\\}

\begin{document}
\maketitle

\section{Introduction}
\begin{figure}[h]
\begin{center}
\setlength{\unitlength}{1in}
\begin{picture}(6.5,3.0)(0,0)
\put(0.48,0.5){\includegraphics[width=2.5in,height=2.6in]{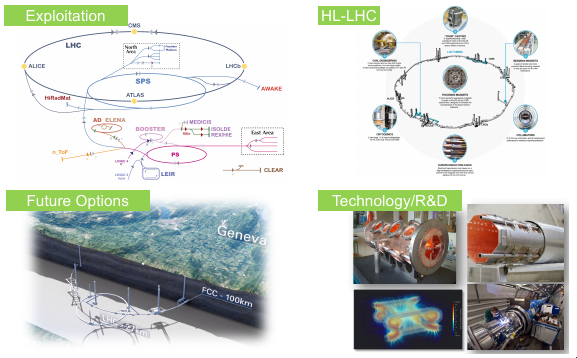}}
\put(3.08,0.5){\includegraphics[width=2.5in,height=2.6in]{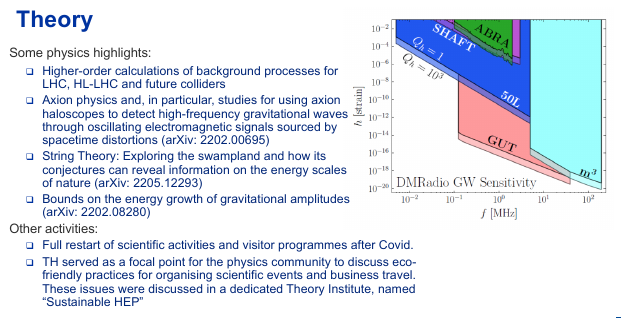}}
\put(1.70,0.25){(a)\hspace{2.40in}(b)}
\end{picture}
\end{center}
\vspace{-10mm}
\caption{\baselineskip=11pt Excerpts from Ref.~\cite{fabiola-1-10-23} on the state of CERN: (a), Future Options and R\& D; (b), Theory highlights.}
\label{fig2}
\end{figure}

\begin{figure}[h!]
\begin{center}
\setlength{\unitlength}{1in}
\begin{picture}(6,2.0)(0,0)
\put(0.70,-0.2){\includegraphics[width=4.5in]{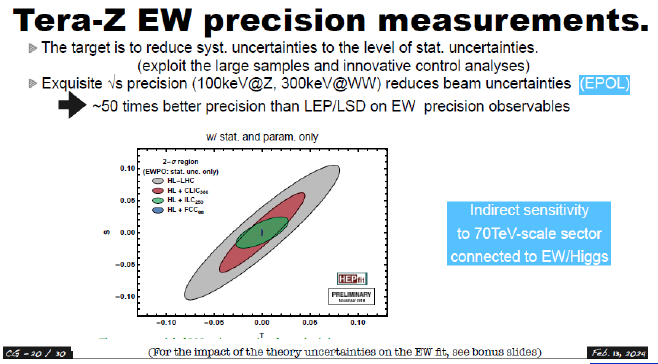}}
\end{picture}
\end{center}
\vspace{-2mm}
\caption{Excerpt from Ref.~\cite{grojean-2-13-24} on expectations for precision EW tests at FCCee.}
\label{fig2a}
\end{figure}

\footnote{This contribution is dedicated in memoriam to my late collaborator Prof. Stanislaw Jadach, who passed away suddenly on Feb. 23, 2023. See his obituary in {\it CERN Courier}, May/June 2023 issue, p.59.}
Future accelerators -- FCC, CLIC, ILC, CEPC, CPPC,  SSC-RESTART, ..., dictate the future of precision theory . If we use FCC as an example, factors of improvement 
from $\sim 5$  to $\sim 100$ are needed from theory. As one can see in Fig.~\ref{fig2} with excerpts from Ref.~\cite{fabiola-1-10-23}, at world-leading laboratories such as CERN the need for precision theory for the success of future collider physics programs is recognized. The figure shows possible future options for CERN featuring the FCC with the important role of higher-order calculations, with their attendant computer algebraic methodology, for its background processes as a theory highlight. We do hope that the funding agencies appreciate the implied connection. Fig.~\ref{fig2a} features an illustration from Ref.~\cite{grojean-2-13-24} which shows the advantage in reach of the FCCee compared to the ILC in precision EW tests of the Standard Theory of elementary particles -- one expects $\sim$ 50 times the precision achieved at LEP/SLD. Resummation is key to such calculations/analyses in many cases. In what follows, we discuss amplitude-based resummation following
the YFS MC methodology~\cite{krkepiph-2024,mttd-2025}.\par
As long as one calculates the corresponding hard radiation residuals to the desired order in the respective coupling, we emphasize that the YFS approach has no limit in principle to its precision~\cite{bardin-zplep1:1989}. This is dissimilar to methods such as the collinear factorization resummation method recently done to subleading log level in Refs.~\cite{frixione-2019,bertone-2019,frixione-2021,bertone-2022} in which various degrees of freedom are integrated out engendering an intrinsic uncertainty. To proceed, after giving a brief recapitulation of the exact amplitude resummation theory in the next Section, we present in Section 4 illustrative results in 
both collider physics and quantum gravity to capture the expanse of the attendant methods.
In Section 5 we discuss improving the collinear limit of YFS theory, one of our last works with Prof. Jadach. Our summary is given in Section 6.\par
We cannot stress too much the following as a key point. Exact amplitude-based resummation realized on an event-by-event basis gives enhanced precision for a given level of exactness: LO, NLO, NNLO, NNNLO, ..., and this is essential for future precision physics as exemplified by CERN -- computer algebraic methods~\cite{blum-schneider-2023,bytev-2014} are paramount. \par 
\section{Recapitulation of YFS Exact Amplitude-Based Resummation}
We include here a synopsis of exact amplitude-based resummation theory, as it is still not generally familiar. The theory is carried by the following master formula:
{\small
\begin{eqnarray}
&d\bar\sigma_{\rm res} = e^{\rm SUM_{IR}(QCED)}
   \sum_{{n,m}=0}^\infty\frac{1}{n!m!}\int\prod_{j_1=1}^n\frac{d^3k_{j_1}}{k_{j_1}} \cr
&\prod_{j_2=1}^m\frac{d^3{k'}_{j_2}}{{k'}_{j_2}}
\int\frac{d^4y}{(2\pi)^4}e^{iy\cdot(p_1+q_1-p_2-q_2-\sum k_{j_1}-\sum {k'}_{j_2})+
D_\rQCED} \cr
&{\tilde{\bar\beta}_{n,m}(k_1,\ldots,k_n;k'_1,\ldots,k'_m)}\frac{d^3p_2}{p_2^{\,0}}\frac{d^3q_2}{q_2^{\,0}},
\label{subp15b}
\end{eqnarray}}
where the {\em new}\footnote{The {\em non-Abelian} nature of QCD requires a new treatment of the corresponding part of the IR limit~\cite{Gatheral:1983} so that we usually include in ${\rm SUM_{IR}(QCED)}$ only the leading term from the QCD exponent in Ref.~\cite{Gatheral:1983} -- the remainder is included in the residuals $\tilde{\bar\beta}_{n,m}$ .}(YFS-style) residuals   
{$\tilde{\bar\beta}_{n,m}(k_1,\ldots,k_n;k'_1,\ldots,k'_m)$} have {$n$} hard gluons and {$m$} hard photons. The new residuals and the  infrared functions ${\rm SUM_{IR}(QCED)}$ and ${ D_\rQCED}$ are defined in Ref.~\cite{mcnlo-hwiri,mcnlo-hwiri1}.  For illustration, we note that, to meet the precision tags required for the FCCee, we will need the $\tilde{\bar\beta}_{n,m}$ exact in ${\cal O}\left(\frac{\alpha}{\pi}, \frac{\alpha}{\pi}L, \left(\frac{\alpha}{\pi}\right)^2, \left(\frac{\alpha}{\pi}\right)^2L, \left(\frac{\alpha}{\pi}\right)^2L^2, \left(\frac{\alpha}{\pi}\right)^3L^2, \left(\frac{\alpha}{\pi}\right)^3L^3,
 \left(\frac{\alpha}{\pi}\right)^4L^4\right)$ where $L = \ln\frac{s}{m_e^2}$ is the usual big log in an obvious notation. This need will be satisfied by using computer-algebraic methods to evaluate the attendant implied Feynman diagrams.  Parton shower/ME matching engenders, as explained in Ref.~\cite{mcnlo-hwiri,mcnlo-hwiri1}, the replacements {$\tilde{\bar\beta}_{n,m}\rightarrow \hat{\tilde{\bar\beta}}_{n,m}$} and thereby allows us to connect with MC@NLO~\cite{mcnlo,mcnlo1}, as well as KrkNLO~\cite{krknlo:2015}, via the basic formula{\small
\begin{equation}
{d\sigma} =\sum_{i,j}\int dx_1dx_2{F_i(x_1)F_j(x_2)} d\hat\sigma_{\rm res}(x_1x_2s).
\label{bscfrla}
\end{equation}}
\par
Eq.(\ref{subp15b}) has paved the way to new results in precision LHC and FCC physics. As an approach to general relativity, one of us (BFLW) has extended the latter equation to general relativity.  New developments accompany our new results in each respective application. We discuss such new results and developments in the next Section.\par
\section{New Developments for Precision Collider Physics: LHC, FCC, CPEC, CPPC, ILC, CLIC, ...} 
\begin{figure}[h!]
\begin{center}
\setlength{\unitlength}{1in}
\begin{picture}(6,2.0)(0,0)
\put(0.5,0){\includegraphics[width=5in]{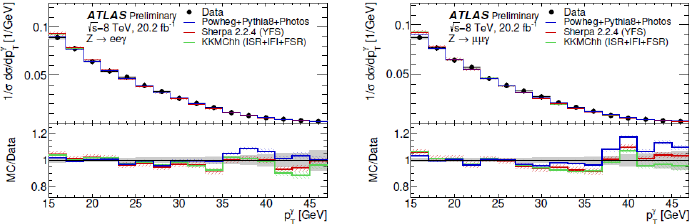}}
\end{picture}
\end{center}
\vspace{-5mm}
\caption{ATLAS analysis of $Z/\gamma$ production at $8$ TeV.}
\label{fig3}
\end{figure}

Four of us (SJ, BFLW, ZAW, SAY) have implemented eq.(\ref{subp15b}) in the MC event generator \KK{MC}-hh~\cite{kkmchh1-sh}. This implementation represents a new development in the expectations for precision physics for the Standard Theory
EW interactions at HL-LHC. This is illustrated by the plots in Fig.~\ref{fig3} in the ATLAS analysis~\cite{atlas-epjc-zg:2024} of $Z\gamma$ production at 8 TeV.
As we see in the figure, in view of the level of uncertainties in the data, the comparison of the data to the Powheg-Pythia8-Photos~\cite{powheg-org,powheg-orga,powheg1,powhega,Sjostrand:2007gs-sh,Golonka:2006tw}, Sherpa2.2.4(YFS)~\cite{sherpa,sherpa-2.2}, and \KK{MC}-hh predictions for the $\gamma p_T$ spectrum shows that, at this point, 
the data are in reasonable agreement with all three predictions. With ~ 10 times the statistics, a precision test against the theories will obtain at HL-LHC.\par
\begin{figure}[h!]
\begin{center}
\setlength{\unitlength}{1in}
\begin{picture}(6,2.0)(0,0)
\put(0.5,0){\includegraphics[width=5in]{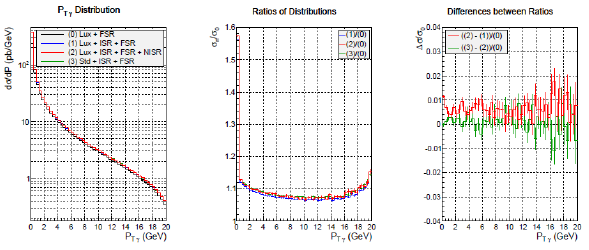}}
\end{picture}
\end{center}
\vspace{-5mm}
\caption{The distribution for $P_{T_\gamma}$ of the photon for which it is greatest for events with at least one photon and each lepton having $ p_{T\ell}> 25 GeV, \eta_\ell< 2.5$ calculated with (0) FSR only (black). (1) FSR + ISR (blue). and (2) FSR + ISR with NISR (red) for NNPDF3.1-LuxQED NLO PDFs. For comparison, (3) shows FSR + ISR with ordinary NNPDF3.1 NLO PDFs (green). The center graph shows
ISR on/off ratios (1)/(0) (blue),(2)/(0) (red) and (3)/(0) (green). The right-hand graph shows the fractional differences ((1)- (2))/(0) in red and ((2)- (3))/(0) in green.}
\label{fig4}
\end{figure}

 
Another new development is the use of Negative ISR (NISR)~\cite{jad-yost-afb,ichep2022-say,sjetaltoappear} evolution to address directly the size of the QED contamination in non-QED PDFs. In a standard notation for PDFs and cross sections, the cross section representation takes the form
\begin{equation}
\begin{split}
\sigma(s)&=
\frac{3}{4}\pi\sigma_0(s)\!\!\!
\sum_{q=u,d,s,c,b} \int d\hat{x}\;  dz dr dt \; \int dx_q dx_\qb\; 
\delta(\hat{x}-x_qx_\qb z t)
\\&\times
f^{h_1}_q(   s\hat{x}, x_q) 
f^{h_2}_\qb( s\hat{x}, x_\qb) \;
 \rho_I^{(0)}\big(\gamma_{Iq}(s\hat{x}/m_q^2),z\big)\; 
 \rho_I^{(2)}\big(-\gamma_{Iq}(Q_0^2/m_q^2),t\big)\; 
\\&\times
\sigma^{Born}_{q\qb}(s\hat{x}z)\;
\langle W_{MC} \rangle,
\label{eq:kkhhsigmaPru}
\end{split}
\end{equation}
which includes an extra convolution with the well known second order exponentiated ISR
``radiator function'' $\rho_I^{(2)} $ with the negative evolution time argument
$-\gamma_{Iq}(Q_0^2/m_q^2)$ defined in Ref.~\cite{jad-yost-afb}. This removes the QED below $Q_0$. We illustrate this in Fig.~\ref{fig4} from Ref.~\cite{ichep2022-say} 
for the $P_{T_\gamma}$ for the photon for which it is the largest in $Z\gamma^*$ production and decay to lepton pairs at $8 ~TeV$ at the LHC at for
cuts as described in the figure. The results in the figure show, in agreement with arguments in Ref.~\cite{kkmchh2}, that the effect of QED contamination in non-QED
PDFs is below the errors on the PDFs.\par
\begin{figure}[h]
\begin{center}
\setlength{\unitlength}{1in}
\begin{picture}(6.5,3.0)(0,0)
\put(0,0.5){\includegraphics[width=3.0in,height=1.5in]{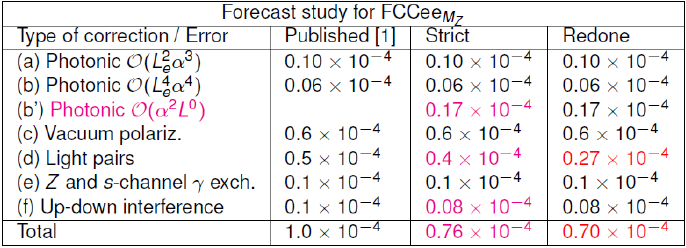}}
\put(3.05,0.5){\includegraphics[width=3.0in,height=1.5in]{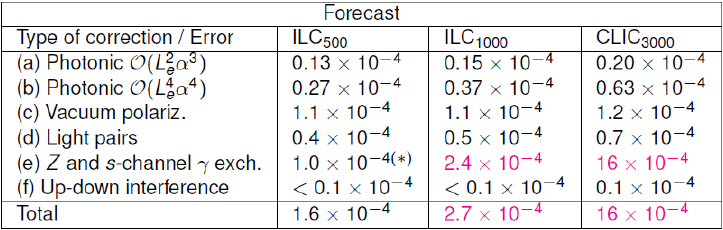}}
\put(1.5,0.25){(a)\hspace{2.8in}(b)}
\end{picture}
\end{center}
\vspace{-10mm}
\caption{\baselineskip=11pt Current purview on luminosity theory errors: (a), FCC-ee at $M_Z$; (b), proposed higher energy colliders}
\label{fig5}
\end{figure}

Five of us (SJ, WP, MS, BFLW, SAY), in view of the planned EW/Higgs factories, have discussed in Refs.~\cite{Jadach:2018jjo,Jadach:2021ayv-sh,fcc2023wkshpms,skrzypek-2023mttd,placzek-2025mttd} the new developments for the BHLUMI~\cite{bhlumi4:1996} luminosity theory error. These new developments are illustrated in Fig.~\ref{fig5}~\cite{fcc2023wkshpms,skrzypek-2023mttd} wherein we show the current purview for the FCC-ee at $M_Z$ and that for the proposed higher energy colliders. 
In addition to the improvements at $M_Z$ shown in Fig.~\ref{fig5}(a) to 0.007\%, there is the possibility that item (c) in Fig~\ref{fig5}(a) could be reduced by a factor of ~ 6 by the use of the results in Ref.~\cite{fjeger-fccwksp2019} together with lattice methods~\cite{latt1,latt2,latt3}. The formula to be studied is $$\Delta\alpha_{had}(t)=\Delta\alpha_{had}(-Q^2_0)|_{lat}+[\Delta\alpha_{had}(t)-\Delta\alpha_{had}(-Q^2_0)]|_{pQCDAdler}$$ with {\it lat} denoting the methods of Refs.~\cite{latt1,latt2,latt3} and {\it pQCDAdler} denoting the methods of Ref.~\cite{fjeger-fccwksp2019}.  In this latter connection, see also \cite{fjeger2025}.\par
One of us (BFLW) has shown in Refs.~\cite{ward:2013dkunv,ijmpa2018} that the UV divergences of quantum gravity are tammed by amplitude-based resummation. There are many consequences, one of which we illustrate here as follows, using a standard notation,  
\begin{equation}
\begin{split}
\rho_\Lambda(t_0)&\cong \frac{-M_{Pl}^4(1+c_{2,eff}k_{tr}^2/(360\pi M_{Pl}^2))^2}{64}\sum_j\frac{(-1)^Fn_j}{\rho_j^2}\cr
          &\qquad\quad \times \frac{t_{tr}^2}{t_{eq}^2} \times (\frac{t_{eq}^{2/3}}{t_0^{2/3}})^3\cr
    &\cong \frac{-M_{Pl}^2(1.0362)^2(-9.194\times 10^{-3})}{64}\frac{(25)^2}{t_0^2}\cr
   &\cong (2.4\times 10^{-3}eV)^4.
\end{split}
\label{eq-rho-expt}
\end{equation}
Here, the age of the universe is $t_0$ and we take it to be $t_0\cong 13.7\times 10^9$ yrs, the transition time between the Planck regime and the classical Friedmann-Robertson-Walker(FRW) regime is $t_{tr}\sim 25 t_{Pl}$~\cite{reuter2,ward:2013dkunv,ijmpa2018}, and the vacuum index~\cite{ward:2013dkunv,ijmpa2018} of the ST is $$c_{2,eff}\cong 2.56 \times 10^4.$$ In the estimate in (\ref{eq-rho-expt}), the first factor in the second line comes from the radiation dominated period from
$t_{tr}$ to $t_{eq}$ engenders the first factor and the matter dominated period from $t_{eq}$ to $t_0$ engenders the second factor. The experimental result~\cite{pdg2008}\footnote{See also Ref.~\cite{sola2,sola3} for analyses that suggest 
a value for $\rho_\Lambda(t_0)$ that is qualitatively similar to this experimental result.} 
$\rho_\Lambda(t_0)|_{\text{expt}}\cong ((2.37\pm 0.05)\times 10^{-3}eV)^4$ is close to the estimate in (\ref{eq-rho-expt}). 
\par
\begin{figure}[H]
\begin{center}
\setlength{\unitlength}{1in}
\begin{picture}(6,1.56)(0,0)
\put(1.5,1.55){\bf Improved Weight Distribution for $\nu_e$ Events}
\put(0.5,0){\includegraphics[width=5in]{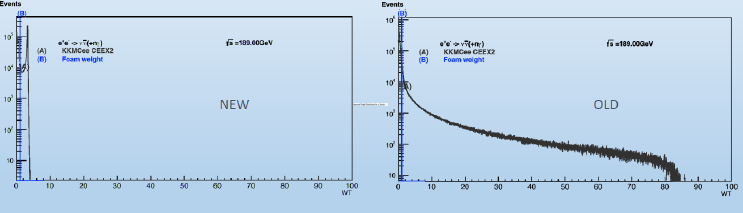}}
\end{picture}
\end{center}
\vspace{-5mm}
\caption{The black curves (A) are the CEEX ${\cal O}(\alpha^2)$ weights. FOAM's internal weight was set to 1 (B).}
\label{figwgt-dist}
\end{figure}

One of ur more important recent developments is the release of KKMCee--v 5.00~\cite{Jadach:2022vf,ichep2024-say}.
The program has been transcoded from Fortran to C++. In addition to the upgraded version of DIZET~\cite{Arbuzov:2020coe}
for its EW library, which is now run in advance to create EW tables, the code also includes various models of beam energy spread for FCC and ILC/CLIC. A new analytical tool KKeeFoam is available for cross-checks with analytical resummation including IFI. Thanks to the introduction of auxillary C++ classses, the most advanced CEEX QED matrix element is now more compact and transparent. Its HEPMC3 event record facilitates various interfaces such as that to modern parton showers. See Refs.~\cite{Jadach:2022vf,ichep2024-say} for more detailed discussion of the improvements carried by v.--5.00, not the least of which is that the underlying MC algorithm has been replaced by a general-purpose adaptive MC generator FOAM~\cite{Jadachfm:1999}. Here, we illustrate these improvements with the weight distribution in Fig.~\ref{figwgt-dist}. The effect is most dramatic for $\nu_e$ generation above 105 GeV, 
\noindent where a strong peak at $\theta_f = 0$ begins to form due to
$t$-channel W exchange. The graphs in the figure compare the same weight distributions for the FOAM-based algorithm (NEW in the figure) and the FORTRAN algorithm (OLD in the figure)  for $e^+e^-\rightarrow \bar{\nu}_e\nu_e+n\gamma$ at $\sqrt{s} =189$ GeV. There is a factor of 20 improvement.\par
\section{Improved Collinear Limit in YFS Theory}
Improving the collinear limit of YFS theory is motivated by the higher precision requirements for HL-LHC/FCC physics. In the usual YFS theory the virtual infrared function $B$ in the s-channel resums ~\cite{Jadach:2023cl1}  (exponentiates) the non-infrared term
$\frac{1}{2}Q_e^2{\alpha\over\pi} L$ in $e^+(p_2)\;e^-(p_1)\rightarrow \bar{f}(p_4)\;e^-(p_3)$ using an obvious notation where the respective big log is $L = \ln(s/m_e^2)$ when $s=(p_1+p_2)^2$ is the center-of-mass energy squared. It is also shown in Ref.~\cite{gribv-lptv:1972} that the term $\frac{3}{2}Q_e^2{\alpha\over\pi} L$ exponentiates -- see also Refs.~\cite{frixione-2019,bertone-2019,frixione-2021,bertone-2022} for recent developments in the
attendant collinear factorization approach. Does the YFS theory allow an extension that would also exponentiate the latter term? Three of us (SJ, BFLW, ZAW) have answered this question in the affirmative in Ref.~\cite{Jadach:2023cl1}.\par
Specifically, the virtual infrared function $B$ in the s-channel can be extended~\cite{Jadach:2023cl1} to
\begin{equation}
\begin{split}
B_{CL}&\equiv B+{\bf \Delta B}\\
          &= \int {d^4k\over k^2} {i\over (2\pi)^3} 
                       \bigg[\bigg( {2p-k \over 2kp-k^2} - {2q+k \over 2kq+k^2} \bigg)^2{\bf -\frac{4pk-4qk}{(2pk-k^2)(2qk+k^2)}}\bigg],
\end{split}
\end{equation}
while the real infrared function $\tilde{B}$ can be extended~\cite{Jadach:2023cl1} to 
\begin{equation}
\begin{split}
\tilde{B}_{CL} &\equiv \tilde{B}+{\bf \Delta\tilde{B}}\\
 &=\frac{-1}{8\pi^2}\int\frac{d^3k}{k_0}\bigg\{\large(\frac{p_1}{kp_1} - \frac{p_2}{kp_2}\large)^2 +{\bf \frac{1}{kp_1}\large(2 -\frac{kp_2}{p_1p_2}\large)}\\
                                          &\qquad\qquad+{\bf \frac{1}{kp_2}\large(2 -\frac{kp_1}{p_1p_2}\large)}\bigg\},
\end{split}
\label{eq-real2}
\end{equation}
where the extensions are indicated in boldface in an obvious notation. The YFS infrared algebra is unaffected by these extensions while the $B_{CL}$ does exponentiate the entire $\frac{3}{2}Q_e^2{\alpha\over\pi} L$ term and the $\tilde{B}_{CL}$ does carry the respective collinear big log of the exact result in Ref.~\cite{berends-neerver-burgers:1988} in the soft regime.\par 
For CEEX, the corresponding collinear extension of the soft eikonal amplitude factor defined in Ref.~\cite{ceex2:1999sh}
for the photon polarization $\sigma$ and $e^-$  helicity $\sigma'$ is given by 
\begin{equation}
\begin{split}
\sfac_{CL,\sigma}(k) = \sqrt{2}Q_ee\bigg[-\sqrt{\frac{p_1\zeta}{k\zeta}}\frac{<k\sigma|\hat{p}_1 -\sigma>}{2p_1k}
    +{\bf \delta_{\sigma'\;-\sigma}\sqrt{\frac{k\zeta}{p_1\zeta}}\frac{<k\sigma|\hat{p}_1 \sigma'>}{2p_1k}}\\
    + \sqrt{\frac{p_2\zeta}{k\zeta}}\frac{<k\sigma|\hat{p}_2 -\sigma>}{2p_2k}+{\bf \delta_{\sigma' \sigma}\sqrt{\frac{k\zeta}{p_2\zeta}}\frac{<\hat{p}_2 \sigma'|k -\sigma>}{2p_2k}}\bigg],
\end{split}
\label{eq-real4}
\end{equation}
where from Ref.~\cite{ceex2:1999sh} $\zeta\equiv (1,1,0,0)$ for our choice for the respective auxiliary vector in our Global Positioning of Spin (GPS)~\cite{gps:1998} spinor conventions with the consequent definition $\hat{p}= p - \zeta m^2/(2\zeta p)$
for any four vector $p$ with $p^2 = m^2.$ The collinear extension terms are again indicated in boldface.\par
We expect these extended infrared functions to give in general a higher precision for a given level of exactness~\cite{bflwetaltoappear}.\par
\section{SUMMARY}
Amplitude-based resummation allows improved control of both the IR and Collinear limits. The attendant MC realizations are needed for present and future precision physics. Our new collinearly enhanced soft functions imply a higher level of accuracy for a given level of exactness in the IR-finite YFS hard photon residuals. We have an enhanced toolbox which is available to extend the (CEEX) YFS MC method to other important processes at present and future colliders -- some New Physics may hang in the balance.\par
\section*{Acknowledgements}
W. P{\l}aczek, A. Siodmok, M. Skrzypek, and Z. Was were supported in part by Narodowe Centrum Nauki, Poland, grant No. 2023/50/A/ST2/00224 and S. Yost was supported in part by a grant from The Citadel Foundation.  We thank Prof. Gian Giudice for the support and kind hospitality of the CERN TH Department while some of this work was in progress.\par
\baselineskip=14pt
\bibliography{Tauola_interface_design}{}
\bibliographystyle{utphys_spires}

\end{document}